\documentclass[aps,10pt,pra,reprint,floatfix,superscriptaddress]{revtex4-1}
\usepackage{graphicx,textcomp,dsfont,hyperref,natbib,color,amsmath,ifpdf}
\pdfoutput=1

\hypersetup{
  colorlinks   = true, 
  urlcolor     = blue, 
  linkcolor    = black, 
  citecolor   = blue 
}

\begin{document}
\title{Composite pulses for interferometry in a thermal cold atom cloud}
\author{Alexander Dunning}
\author{Rachel Gregory}
\author{James Bateman}
\author{Nathan Cooper}
\author{Matthew Himsworth}
\affiliation{School of Physics \& Astronomy, University of Southampton, Highfield, Southampton SO17 1BJ, UK}
\author{Jonathan A. Jones}
\affiliation{Centre for Quantum Computation, Clarendon Laboratory, University of Oxford, OX1 3PU, UK}
\author{Tim Freegarde}
\affiliation{School of Physics \& Astronomy, University of Southampton, Highfield, Southampton SO17 1BJ, UK}
\date{\today}


\begin{abstract}
Atom interferometric sensors and quantum information processors must maintain coherence while the evolving quantum wavefunction is split, transformed and recombined, but suffer from experimental inhomogeneities and uncertainties in the speeds and paths of these operations. Several error-correction techniques have been proposed to isolate the variable of interest. Here we apply composite pulse methods to velocity-sensitive Raman state manipulation in a freely-expanding thermal atom cloud. We compare several established pulse sequences, and follow the state evolution within them. The agreement between measurements and simple predictions shows the underlying coherence of the atom ensemble, and the inversion infidelity in a $\sim 80\, \mu{\mathrm K}$ atom cloud is halved. Composite pulse techniques, especially if tailored for atom interferometric applications, should allow greater interferometer areas, larger atomic samples and longer interaction times, and hence improve the sensitivity of quantum technologies from inertial sensing and clocks to quantum information processors and tests of fundamental physics.

\end{abstract}
\maketitle

\section{INTRODUCTION}
Emerging quantum technologies such as atom interferometric sensors~\cite{McGuirk2002Sensitive}, fountain atomic clocks~\cite{Wynands2005} and quantum information processors~\cite{Schmidt-Kaler2003a} rely upon the precise manipulation of quantum state superpositions, and require coherence to be maintained with high fidelity throughout extended sequences of operations that split, transform and recombine the wavefunction. In practice, however, inhomogeneities lead to uncertainty in the rates and phase space trajectories of these operations. To reduce the sensitivity of the intended operation to variations in laser intensity, atomic velocity, or even gravitational acceleration~\cite{Bertoldi2010}, several approaches have been proposed, from quantum error correction~\cite{Steane1996} to shaped pulses~\cite{Torosov2011a} and rapid adiabatic passage~\cite{Baum1985,Bateman2007Fractional,Kovachy2012Adiabaticrapidpassage}. Just as squeezing does for an individual wavefunction~\cite{*[{See e.g. }][{}] Horrom2012}, these techniques aim to reduce the uncertainty projected within an ensemble distribution upon the parameter of interest.

NMR spectroscopists have over many years developed `composite pulse' techniques to compensate for systematic variations in the speed and trajectory of coherent operations, and thus refocus a quantum superposition into the desired state~\cite{Levitt1981,Levitt1986,Cummins2003,Vandersypen2004}. The various pulse sequences differ in their tolerance of `pulse length' (or coupling strength) and `off-resonance' errors and correlations between them, and in the operations for which they are suitable and the properties whose fidelity they protect. All are in principle applicable to the coherent control of any other two-state superposition, and such techniques have been applied to the manipulation of superconducting qubits~\cite{Collin2004}, diamond NV colour-centres~\cite{Aiello2012}, trapped ions~\cite{Gulde2003,Schmidt-Kaler2003a,Chiaverini2004,Riebe2007,Timoney2008,Huntemann2012}, microwave control of neutral atoms~\cite{Hart2007,Rakreungdet2009Accurate,Lundblad2009,Leroux2009,Schleier-Smith2010}, and even the polarization of light~\cite{Dimova2013}.

Perhaps the simplest composite pulse sequence, based upon Hahn's spin-echo~\cite{Hahn1950}, inserts a phase-space rotation between two halves of an inverting `$\pi$-pulse' to compensate for systematic variations in the coupling strength or inter-pulse precession rate. A number of researchers have applied such schemes to optical pulses in atom interferometry, using the $\pi$-pulse also to ensure proper path overlap analogous to the mirrors of a Mach--Zehnder interferometer. Using stimulated Raman transitions from a single Zeeman substate in a velocity-selected sample of cold Cs atoms, Butts {\em et al.}~\cite{Butts2013Efficient} extended this scheme by replacing the second $\pi/2$-pulse with one three times as long, thus forming a \textsc{waltz} composite pulse sequence~\cite{Shaka1983Improved} that, with appropriate optical phases, is tolerant of detuning errors and hence the Doppler broadening of a thermal sample. Following the proposal of McGuirk {\em et al.}~\cite{McGuirk2002} that composite pulses could withstand the Doppler and field inhomogeneities in `large-area' atom interferometers, in which additional $\pi$-pulses increase the enclosed phase space area to raise the interferometer sensitivity, Butts {\em et al.} showed that the \textsc{waltz} pulse increased the fidelity of such augmentation pulses by around 50\%.

In this paper, we use velocity-sensitive stimulated Raman transitions to compare the effectiveness of several established pulse sequences upon an unconfined sample of \textsuperscript{85}Rb atoms, distributed across a range of Zeeman substates, after release from a magneto-optical trap. We explore the \textsc{corpse}~\cite{Cummins2000Use}, \textsc{bb1}~\cite{Wimperis1994Broadband}, \textsc{knill}~\cite{Ryan2010Robust} and \textsc{waltz}~\cite{Shaka1983Improved} sequences, determine both the detuning dependence and the temporal evolution in each case, and show that the inversion infidelity in a $\sim 80\, \mu{\mathrm K}$ sample may be halved from that with a basic `square' $\pi$-pulse. Comparison with simple theoretical predictions shows the underlying coherence of the atomic sample, and suggests that if cooled towards the recoil limit such atoms could achieve inversion fidelities above 99\%. Our results demonstrate the feasibility of composite pulses for improving pulse fidelity in large-area atom interferometers and encourage the development of improved pulse sequences that are tailored to these atomic systems~\cite{Jones2013}; they also open the way to interferometry-based optical cooling schemes such as those proposed in \cite{Weitz2000} and \cite{Freegarde2003}.

\section{EXPERIMENT}
\begin{figure}[b!]
  \centering
  \includegraphics[width=8.5cm]{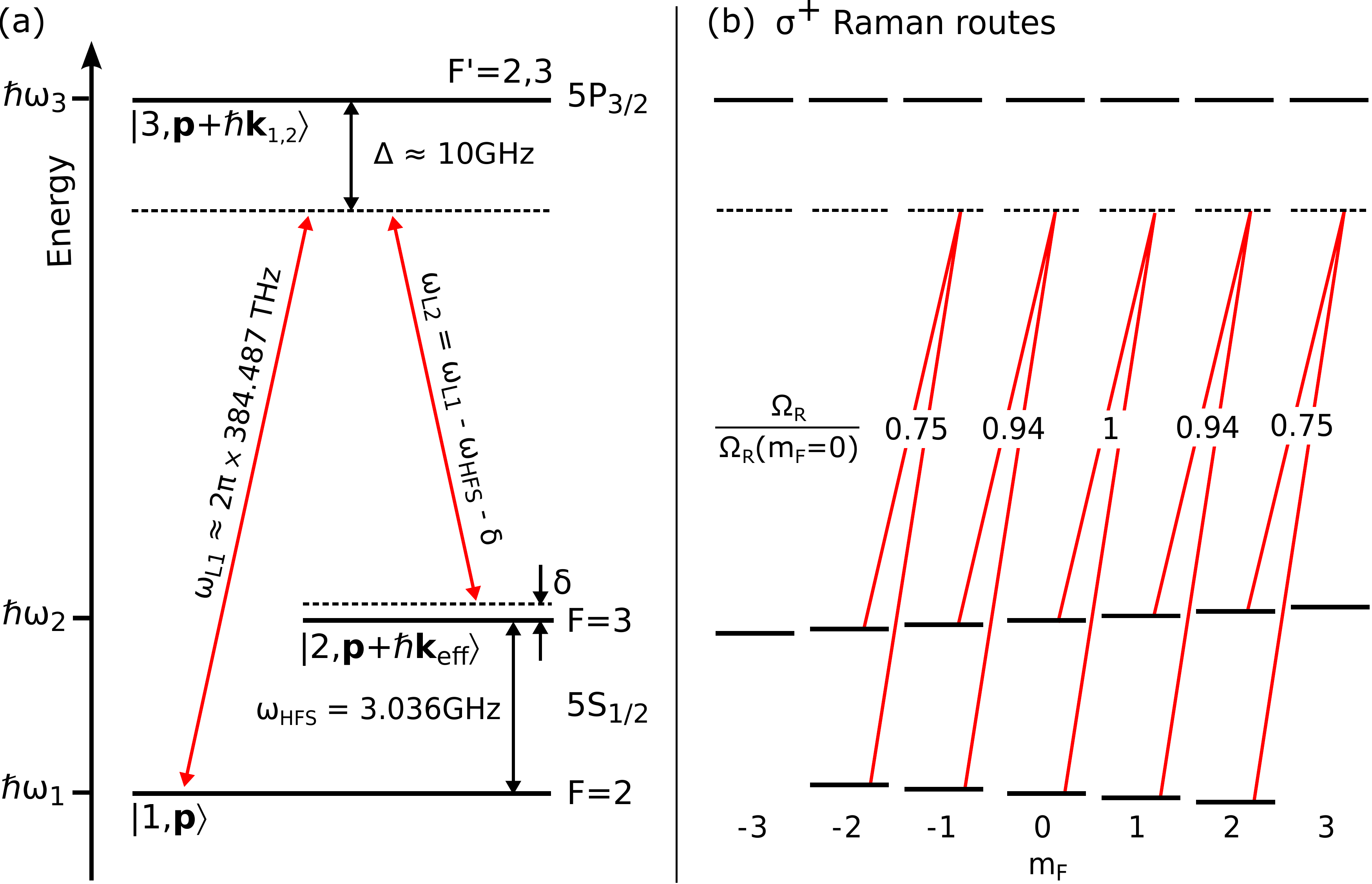}
  \caption{\label{RamanDiagramPaper}Energy levels (a) for stimulated Raman transitions in $^{85}$Rb, and (b) individual dipole-allowed Raman routes for atoms initialised in the different Zeeman $m_F$ sub-levels, where the counterpropagating Raman beams are opposite-circular polarised $\sigma^{+} - \sigma^{-}$. The relative transition strengths, calculated from the Clebsch-Gordan coefficients, normalised to the 0--0 transition, are given for each route.}
  \label{RamanDiagram1}
\end{figure}

We explore a popular atom interferometer scheme, used to measure gravitational acceleration~\cite{Kasevich1992Measurement,McGuirk2002Sensitive}, rotation~\cite{Riedl2013} and the fine-structure constant~\cite{Weiss1994Precision}, in which stimulated Raman transitions~\cite{Kasevich1991Atomic} between ground hyperfine states provide the coherent `beamsplitters' and `mirrors' to split, invert and recombine the atomic wavepackets; motion, acceleration or external fields then induce phase shifts between the interferometer paths that are imprinted on the interference pattern at the interferometer output. Our experiments are performed on a cloud of about $2 \times 10^{7}$ \textsuperscript{85}Rb atoms with a temperature of $50$--$100\, \mu$K, and the 780~nm Raman transition is driven between the $F=2$ and $F=3$ ground states (Figure~\ref{RamanDiagram1}(a)). The two laser fields are detuned ($\Delta$) from single-photon resonance by many GHz to avoid population of, and spontaneous emission from, the $5P_{3/2}$ intermediate state. Following extinction of the lasers and magnetic fields of the conventional magneto-optical trap, the atoms are prepared in the $5S_{1/2},\,F=2$ state in a distribution across the five Zeeman states $m_{F} = -2 \ldots +2$, which with the magnetic field off are degenerate to within $5$~kHz.

\begin{figure}[t!]
  \centering
    \includegraphics[width=8.4cm]{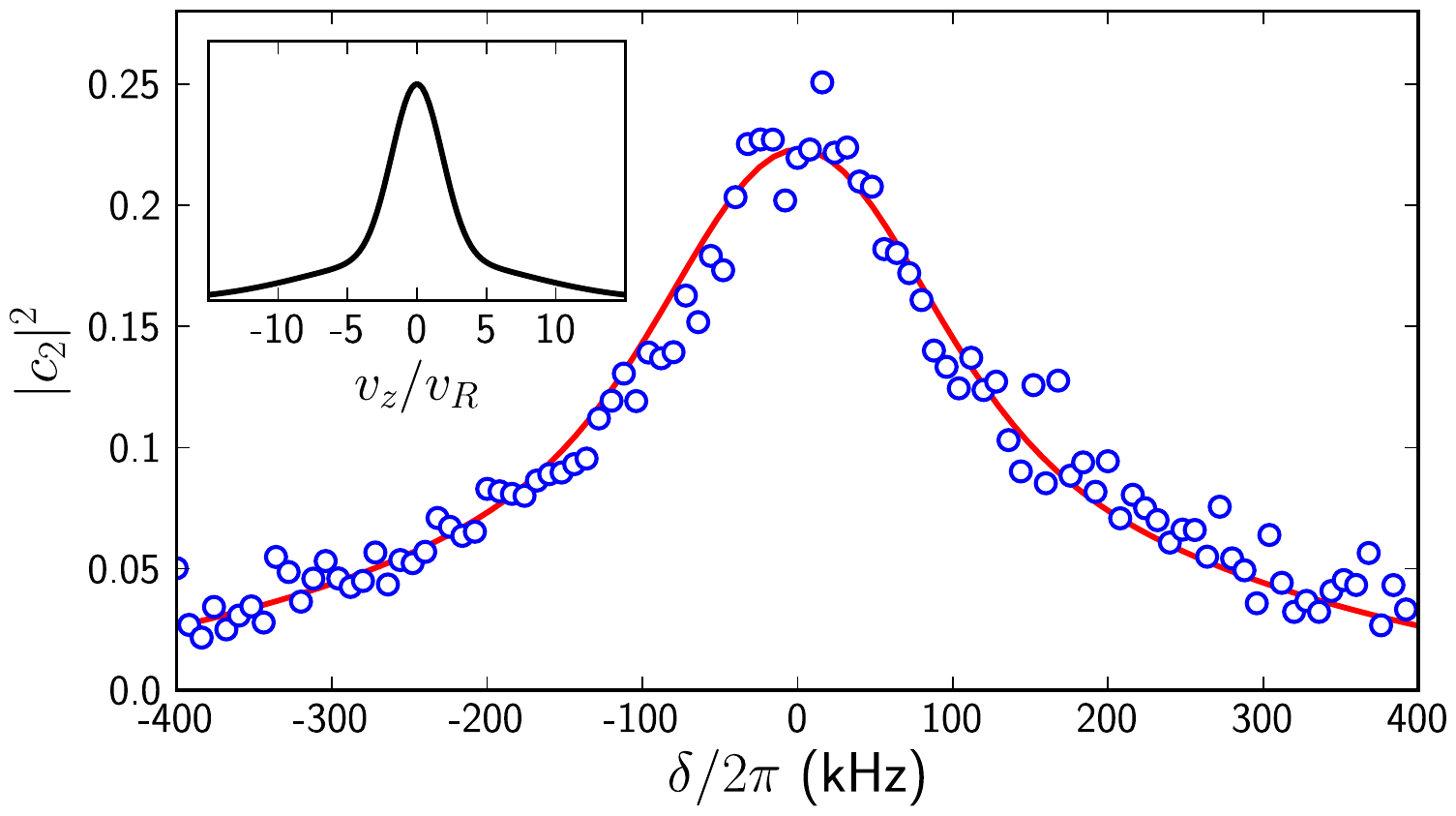}
    \caption{Doppler-broadened Raman lineshape~\cite{Reichel1994Subrecoil} after the molasses phase. Measured data (circles) match a simulation (dashed) for a double-Gaussian with similar populations in the central peak ($4.8\,\mu$K) and broader background ($83\,\mu$K). $\delta$ 
    is the Raman detuning from the light-shifted hyperfine splitting. (Inset) Deduced atom cloud velocity distribution; $v_R$ is the Raman recoil velocity. The mean temperature is $45\,\mu$K.}
  \label{VeloDist}
\end{figure}

We use counterpropagating Raman beams, which impart twice the  impulse of a single-photon recoil and incur a Doppler velocity dependence that, at low intensities, we use to characterise the atom cloud velocity distribution, as shown in Figure~\ref{VeloDist}. The Raman pulse sequence is then applied and the population of the $F=3$ level is determined by monitoring the fluorescence after pumping to the 5P$_{3/2} \, F=4$ level. If the detuning $\Delta$ is large compared with the $5P_{3/2}$ hyperfine splitting, $\Delta m_{F}=\pm2$ transitions are eliminated and two polarization arrangements are of interest. Opposite-circularly polarized Raman beams drive $\sigma^+$ (or $\sigma^-$) dipole-allowed transitions via the Raman routes shown in Figure~\ref{RamanDiagram1}(b), where, for angular momentum to be conserved, $\Delta m_F = 0$ for the Raman transition regardless of the quantisation axis, but the different coupling strengths lead to different light shifts for different $m_{F}$ sub-states, thus lifting their degeneracy. With orthogonal linear polarizations ($\pi^{+}-\pi^{-}$), which correspond to superpositions of $\sigma^+$ and $\sigma^-$ components, the two $\Delta m_{F}=0$ components add constructively, making the $m_{F}$ dependence of the light shift disappear, and maintaining the degeneracy of the sub-states. For parallel linear polarizations (e.g. $\pi^{+}-\pi^{+}$), the $\Delta m_{F}=0$ components cancel. A more detailed description of the experimental setup and procedures is given in Appendix~\ref{ExpSet}.

The Raman coupling strengths and resonance frequencies depend upon the hyperfine sub-state (shown in Figure~\ref{RamanDiagram1}b), the atom's velocity and, via the light shift, the intensity at the atom's position within the laser beam. These inhomogeneities lead to systematic errors in the manipulation processes and hence dephasing of the interfering components, limiting the interferometric sensitivity.


The effects of experimental inhomogeneities are apparent in Figure~\ref{RotaryEchoes1}a, which shows Rabi flopping in a Zeeman-degenerate atom cloud at $\Omega_{\textnormal{eff}}\approx2\pi\times200$\,kHz, where the mean upper hyperfine state population $|c_2|^2$ is measured as a function of Raman pulse length $t$. The  atoms dephase almost completely within a single Rabi cycle, and the upper state population settles at a transfer fraction of $0.28$. The peak transfer fraction is about $0.5$. The solid curves of Figure~\ref{RotaryEchoes1} are numerical simulations (details given in Appendix~\ref{TheMod}) for equally-populated $m_{F}$ sub-states of the $F=2$ hyperfine state, with uniform illumination and a velocity distribution corresponding to a superposition of two Gaussians as in Figure~\ref{VeloDist}, with parameters given in Table~\ref{SimTable}. Intensity inhomogeneities are included at the observed level of $\sim 7$\% and wash out minor features but contribute little to the overall dephasing.

A common solution~\cite{Butts2013Efficient} to the problem of dephasing is to spin-polarise the atomic ensemble into a single Zeeman sub-state, and pre-select a thermally narrow ($T<1\, \mu$K) portion of its velocity distribution before the Raman pulses are applied. Both of these processes however reduce the atom number and hence the signal-to-noise of the interferometric measurement. Adiabatic rapid passage offers inhomogeneity-tolerant population transfer from a defined initial state, but is inefficient for the recombination of superpositions of arbitrary phase~\cite{Bateman2007Fractional,Weitz1994}. Composite pulses can in contrast operate effectively, in the presence of inhomogeneities, upon a variety of superposition states.

\section{COMPOSITE ROTATIONS FOR ATOM INTERFEROMETRY}
\subsection{Bloch sphere notation}
Coherent operations in atom interferometry may be visualised upon the Bloch sphere, whereby the pure quantum states $\lvert 1\rangle$ and $\lvert 2\rangle$ lie at the poles and all other points on the sphere describe superpositions with various ratios and phases~\cite{Feynman1957}. Raman control field pulses correspond to trajectories of the two-level quantum state vector $\lvert\psi\rangle$ on the surface of the sphere. For constant intensities and frequencies, these are unitary rotations, and the unitary rotation propagator acting on $|\psi\rangle$ takes the form~\cite{Rooney1977}
\begin{eqnarray}
U(\theta,\phi,\alpha) & = & \cos\left(\frac{\theta}{2}\right)\mathds{1} - i\sin\left(\frac{\theta}{2}\right)\big[\mathbf{\sigma_x}\cos(\phi)\cos(\alpha) \nonumber \\
&& +\mathbf{\sigma_y}\sin(\phi)\cos(\alpha)+\mathbf{\sigma_z}\sin(\alpha)\big],
\label{equation1}
\end{eqnarray}
where $\mathbf{\sigma_{x,y,z}}$ are the Pauli spin matrices, and the desired rotation, azimuth and polar angles $\theta$, $\phi$ and $\alpha$ are achieved by setting the interaction time, phase and detuning of the control field respectively. Resonant control fields cause rotations about axes through the Bloch sphere equator ($\alpha = 0$), and result in Rabi oscillations in the state populations as functions of the interaction time; off-resonant fields correspond to inclined axes.

\begin{figure}[t!]
  \centering
    \includegraphics[width=8cm]{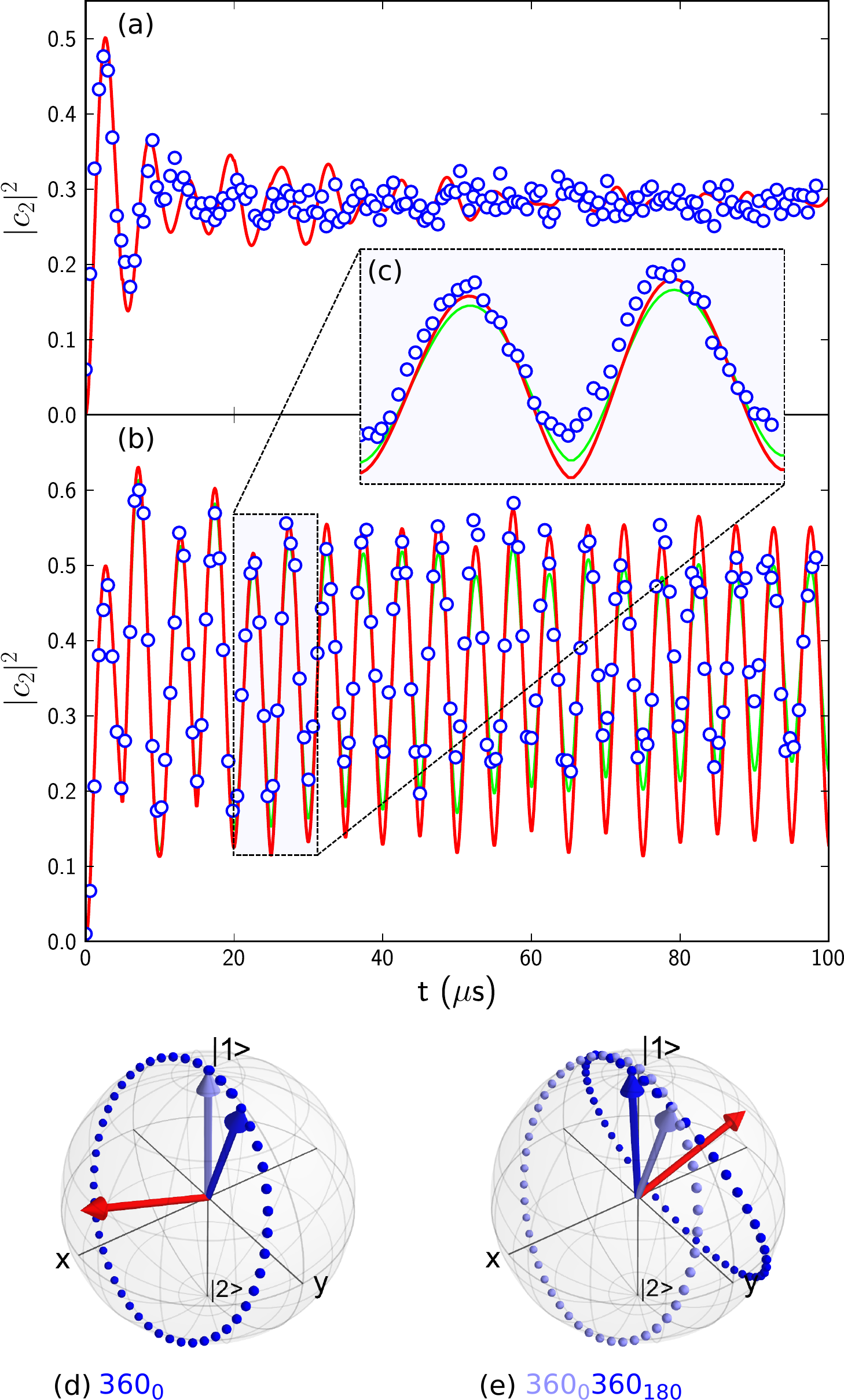}
    \caption{(Color online) Upper hyperfine state population $|c_2|^2$ as a function of Raman interaction time $t$: (a) regular Rabi flopping; (b) Rabi flopping with rotary echoes; (c) highly-sampled data for the indicated portion of (b). Circles are experimental data; lines are numerical simulations, with (green/light-gray) and without (red/dark-gray) phase noise. An example Bloch vector trajectory, starting from $\lvert 1\rangle$ and undergoing the rotary echo $360_{0} 360_{180}$ in the presence of pulse-length and off-resonance errors, is shown (d) before (light-blue/light-grey arrow), during (dots), and at the end of (dark-blue/dark-grey arrow), $360_{0}$; and (e) before, during and after $360_{180}$, at which point the Bloch vector is realigned with $\lvert 1\rangle$. The red/mid-grey arrow is the field vector, around which the Bloch vector rotates.} \label{RotaryEchoes1}
\end{figure}

The pulse sequences explored here all use fields that are set to be resonant for stationary atoms, so we assume $\alpha$ to be zero and write $\theta_{\phi}\equiv U(\theta,\phi,0)$ to represent a rotation defined by the angles $\theta$ and $\phi$ (written in degrees). A sequence of such rotations is written as $\theta_{\phi_1}^{(1)}\theta_{\phi_2}^{(2)}\dots$ where chronological order is from left to right. Two pulses commonly used in atom interferometry are the `mirror' $\pi$ pulse, represented as the rotation $180_{\phi}$, and the `beamsplitter' $\frac{\pi}{2}$ pulse, represented as $90_{\phi}$. On the Bloch sphere, these correspond to half and quarter turns of the state vector about an equatorial axis.

\begin{table*}[t!]
 \begin{ruledtabular}
  \begin{tabular}{l l l c c c c c c c}
  Composite Pulse  & Type &Rotation Sequence $\theta_{\phi}\ldots$	& \multicolumn{2}{l}{Leading order} & total angle &$\mathcal{F}(\sigma^{+}\!-\!\sigma^{+})$ &$\mathcal{F}(\pi^{+}\!-\!\pi^{-})$ \\ \hline
  Rabi $\pi$-pulse    &GR &$180_0$                                         & $\epsilon^2$ & $f^2$ &$180^\circ$	&0.47	& 0.73 \\
  \textsc{corpse}     &GR &$60_{0}300_{180}420_{0}$                        & $\epsilon^2$ & $f^4$ &$780^\circ$	&0.61	& 0.79 \\
  \textsc{knill}      &GR &$180_{240}180_{210}180_{300}180_{210}180_{240}$ & $\epsilon^4$ & $f^4$ &$900^\circ$  &0.64	& \textbf{0.89} \\
  \textsc{bb1}        &GR &$180_{104.5}360_{313.4}180_{104.5}180_0$        & $\epsilon^6$ & $f^2$ &$900^\circ$	&0.56	& 0.80 \\
  90-360-90           &PP &$90_{0}360_{120}90_{0}$                         & $\epsilon^6$ & $f^2$ &$540^\circ$	&0.59	& 0.82 \\
  \textsc{scrofulous} &GR &$180_{60}180_{300}180_{60}$                     & $\epsilon^6$ & $f^2$ &$540^\circ$ 	&0.44	& 0.72 \\
  \textsc{levitt}     &PP &$90_{90}180_{0}90_{90}$                         & $\epsilon^6$ & $f^2$ &$360^\circ$	&0.70	& 0.86 \\
  90-240-90           &GR &$90_{240}240_{330}90_{240}$                     & $\epsilon^2$ & $f^2$ &$420^\circ$	&0.63	& 0.88 \\
  90-225-315          &PP &$90_{0}225_{180}315_{0}$                        & $\epsilon^2$ & $f^2$ &$630^\circ$	&0.71	& \textbf{0.89} \\
  \textsc{waltz}      &PP &$90_{0}180_{180}270_{0}$                        & $\epsilon^2$ & $f^2$ &$540^\circ$	&\textbf{0.77} & 0.88
  \end{tabular}
 \end{ruledtabular}
 \caption{Common composite inversion pulses. The theoretical fidelity $\mathcal{F}$ depends upon the atom cloud temperature as shown in Figure~\ref{VelDistChar}, and is from simulations for typical parameters given in Table~\ref{SimTable}. Bold values indicate best performance at $\delta=0$, which reflects the leading-order terms in the fidelity and their coefficients. PP: point-to-point, GR: general rotor.}
 \label{CPTable}
\end{table*}

\begin{figure*}[t!]
  \centering
  \includegraphics[width=13cm]{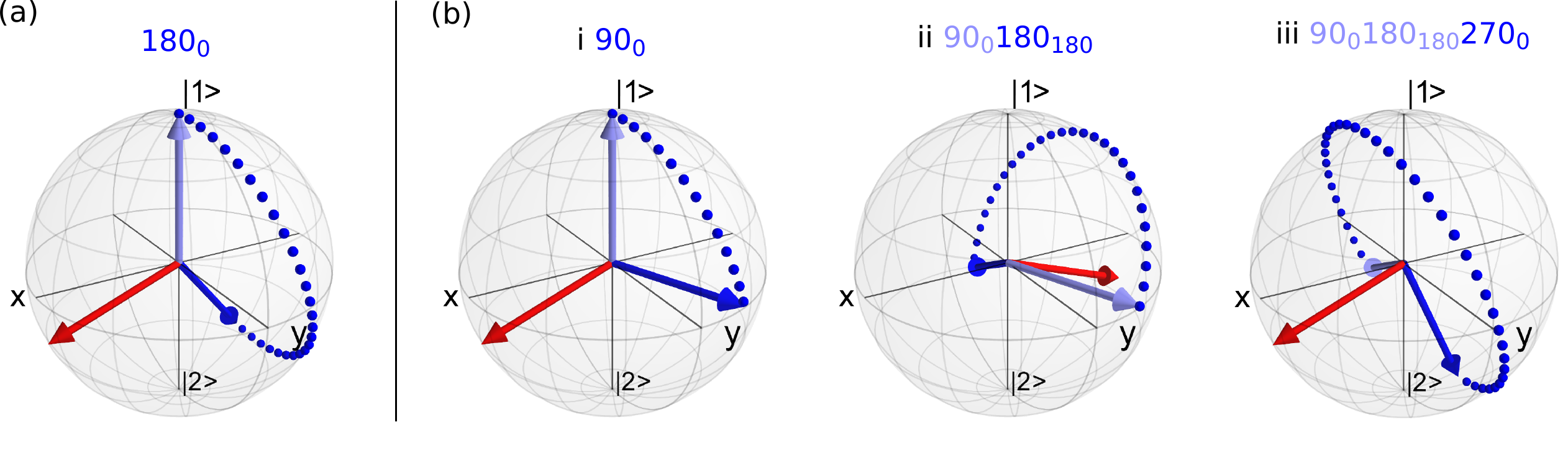}
  \caption{(Color online) Example Bloch sphere trajectories for a Bloch vector starting from $\lvert 1\rangle$, with a Raman detuning $\delta/\Omega_\textnormal{eff} = 0.75$, during (a) a Rabi $\pi$ pulse; and (b) the three contiguous rotations constituting a \textsc{waltz} sequence.
  }
  \label{WaltzBloch}
\end{figure*}

\subsection{Rotary echoes}\label{RabFlo}
A basic means of reducing dephasing in Rabi flopping is the rotary echo~\cite{Rakreungdet2009Accurate,Butts2011}, which may be considered the simplest composite rotation. Reminiscent of Hahn's spin echo~\cite{Hahn1950}, this is a repeated application of the sequence $\theta_{\phi}\theta_{\phi+180}$; when $\theta=360^{\circ}$, as illustrated in Figure~\ref{RotaryEchoes1}d, the $180^{\circ}$ phase shift every whole Rabi cycle causes a periodic reflection of state vector trajectories and realignment, or \emph{echo}, of divergent states. Figure~\ref{RotaryEchoes1}b shows the remarkable reduction in Rabi flopping dephasing obtained with this technique, and the good agreement between experiment and simulation demonstrates the enduring coherence for individual atoms. Simulations for the measured velocity distribution and a Rabi frequency $\Omega_\textnormal{eff} = 2\pi\times200$~kHz (where $t_\pi \equiv \pi/\Omega_{\textnormal{eff}}$ is the pulse duration for optimal ensemble inversion) show flopping with an exponentially falling contrast with a time constant of about $250\,\mu$s, equivalent to 50 Rabi cycles. Experimentally, path length variations, and drifts in the beam intensities and single-photon detuning $\Delta$, cause the fringe visibility to fall over $100-200\,\mu$s; the initial visibility reflects the residual Doppler sensitivity at our modest Rabi frequencies. Inclusion of $1/f$ phase noise (green curve) in the simulation yields closer agreement to the data. The noise level quoted for the modulation electronics is a factor of 3 below that of the simulation, and we therefore expect the major contributor to phase noise at these longer timescales to be path length variation.

\subsection{Composite pulses}\label{ComPul}
As rotary echoes are of limited use beyond revealing underlying coherence, our focus in this paper is upon composite pulses: sequences of rotations that together perform a desired manipulation of the state vector on the Bloch sphere with reduced dephasing from systematic inhomogeneities. Of the many sequences developed for NMR applications~\cite{Merrill2014}, we consider here just a few of interest for inversion in atom interferometry and such experiments. The sequences vary in two key respects.

\begin{figure}[t!]
  \centering
  \includegraphics[width=8cm]{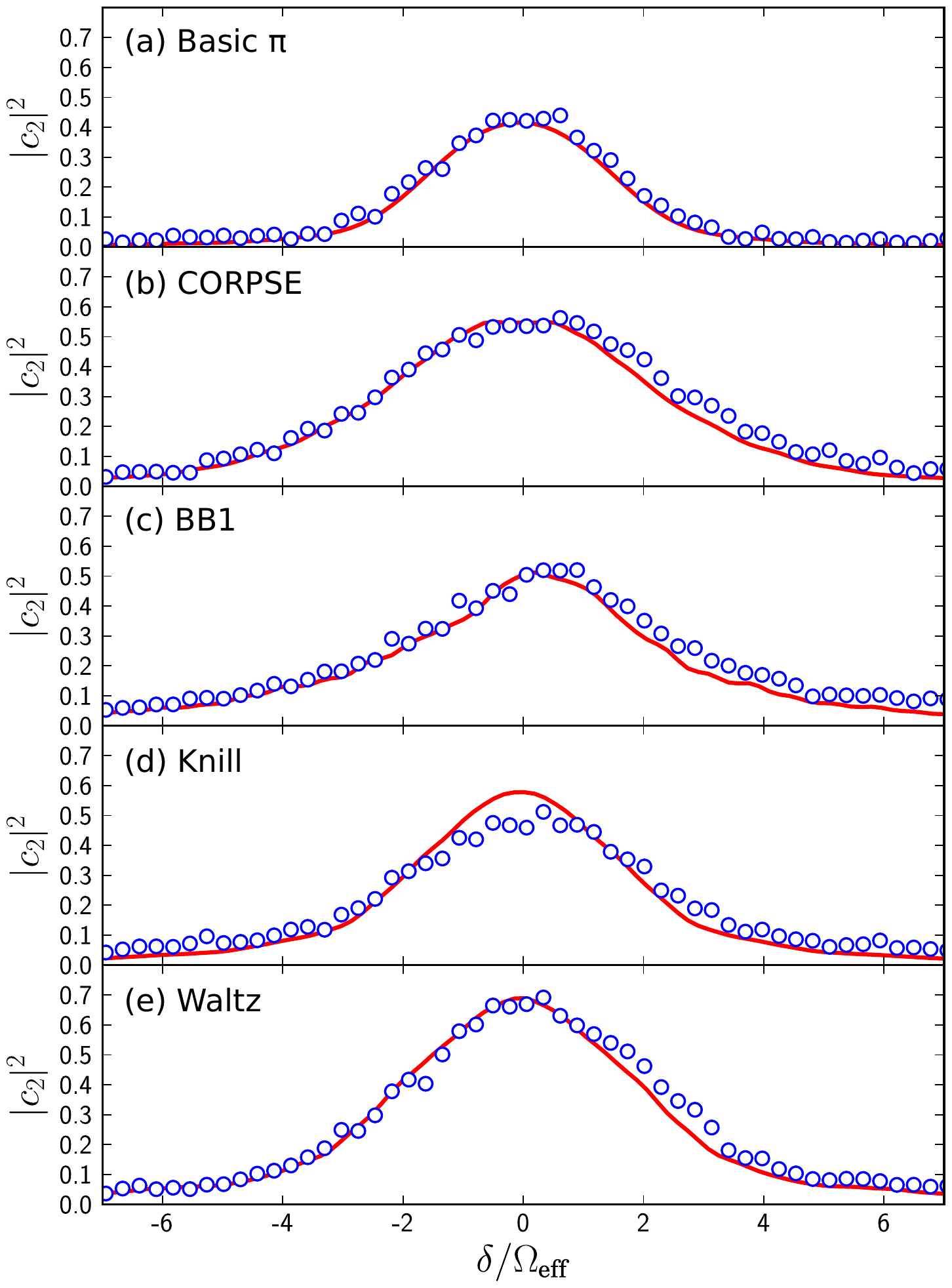}
  \caption{Measured upper state populations (circles) after various $\sigma^{+}-\sigma^{+}$ inversion sequences, as functions of Raman detuning. Simulations (lines) are for a temperature, laser intensity and sublevel splitting found by fitting, within known uncertainties of measured values, to the basic $\pi$ pulse (a) data.
  }
  \label{SpectralScans}
\end{figure}

\begin{figure}[t!]
  \centering
  \includegraphics[width=8cm]{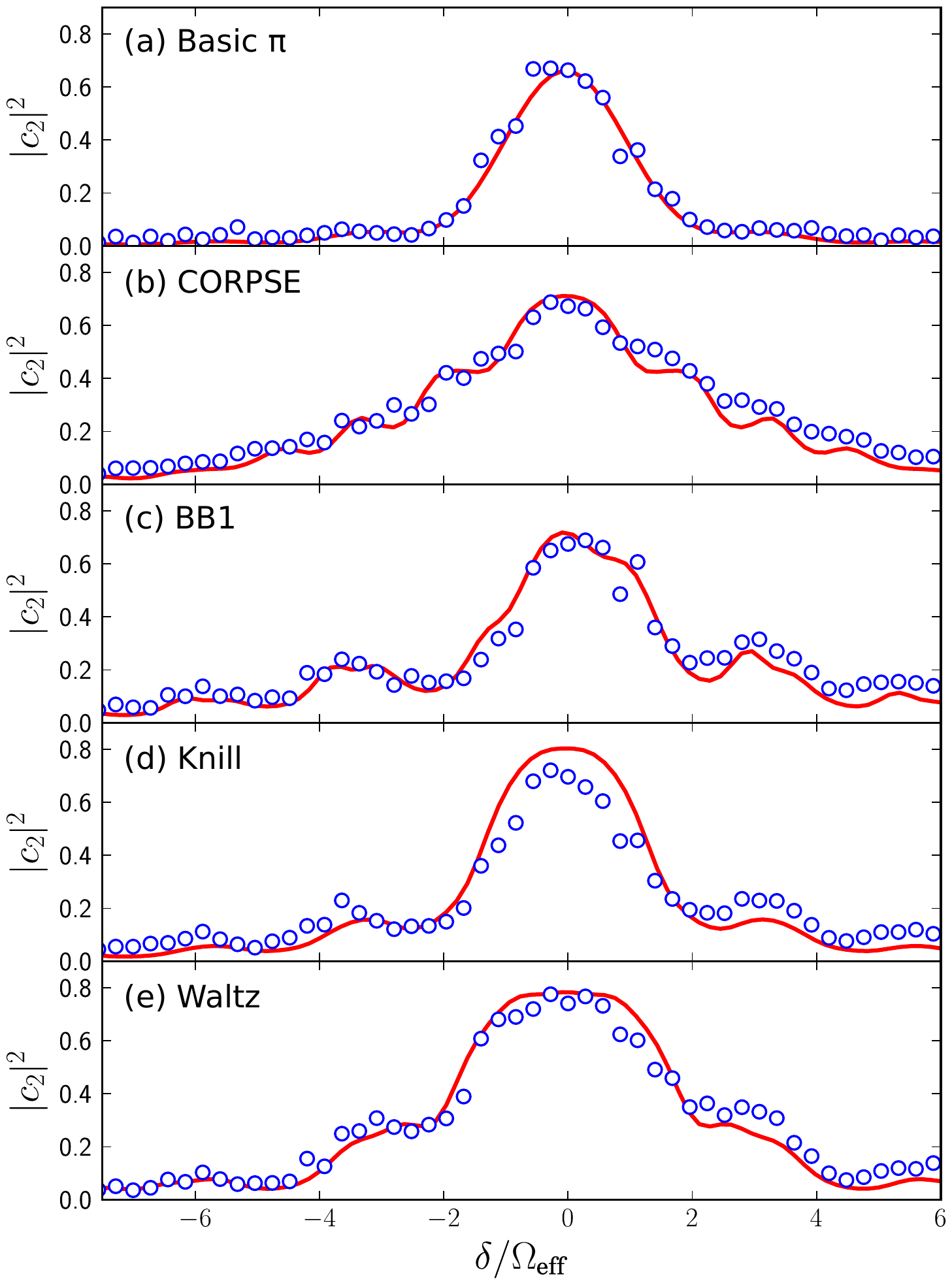}
  \caption{Measured upper state populations (circles) after various $\pi^{+}-\pi^{-}$ inversion sequences, as functions of Raman detuning. Simulations (lines) are for a temperature, laser intensity and sublevel splitting found by fitting, within known uncertainties of measured values, to the basic $\pi$ pulse (a) data.}
  \label{SpectralScansLinLin}
\end{figure}

First, it is common to distinguish between (a) \emph{general rotors}, which are designed to apply the correct unitary rotation to any arbitrary initial state; and (b) \emph{point-to-point} pulses, which work correctly only between certain initial and final states and which for other combinations can be worse than a simple $\pi$~pulse. Some composite inversion pulses suitable for atom interferometry are summarized in Table~\ref{CPTable}.

Secondly, each pulse sequence may be characterized by its sensitivity to variations in the interaction strength and tuning, which instead of the intended rotation propagator $U(\theta,\phi,\alpha)$ result in the erroneous mapping $V(\theta,\phi,\alpha)$. \emph{Pulse-length} (or \emph{-strength}) errors, associated with variations in the strength of the driving field or interaction with it, appear as a fractional deviation $\epsilon=\Delta\theta/\theta$ from the desired rotation angle so that, for the example of a simple Rabi pulse, for small $\epsilon$,
\begin{eqnarray}
V(\theta,0,0) & = & U\big((1+\epsilon)\theta,0,0\big) \nonumber \\
& =  & U(\theta,0,0) \nonumber \\ && - \epsilon\frac{\theta}{2}\left[\sin{\left(\frac{\theta}{2}\right)}\mathds{1} \!+\! i\cos{\left(\frac{\theta}{2}\right)}\mathbf{\sigma_x}\right] \!+\! O(\epsilon^2).\,\,\,\,
\end{eqnarray}
\emph{Off-resonance} errors meanwhile correspond to tilts of the rotation axis due to offsets $f=\delta/\Omega_{\textnormal{eff}}$ in the driving field frequency, so that, for the same example and small $f$,
\begin{eqnarray}
V(\theta,\phi,0) & = & U\left(\theta,\phi,\sin^{-1}(f)\right) \nonumber \\
& = & U(\theta,\phi,0) + f i\sin\left(\frac{\theta}{2}\right)\mathbf{\sigma_z} + O(f^2).
\end{eqnarray}
It is common to describe the dependence upon $\epsilon$ and $f$ of the operation fidelity
\begin{equation}
\mathcal{F} = |\langle\psi|V^{\dagger}U|\psi\rangle|^{2},
\end{equation}
which contains only even powers of $\epsilon$ and $f$. The leading-order uncorrected terms in the corresponding infidelity $\mathcal{I} \equiv 1-\mathcal{F}$ are given in Table~\ref{CPTable}. For Rabi pulses in our system, pulse-length errors are caused by intensity inhomogeneities and mixed transition strengths, and off-resonance errors are due to Doppler shifts. Detunings such as Doppler shifts are also accompanied by high-order pulse-length errors, and intensity variations similarly cause light shifts and thus small off-resonance errors.

We have compared the general rotor sequences known as \textsc{corpse}~\cite{Cummins2000Use}, \textsc{bb1}~\cite{Wimperis1994Broadband} and \textsc{knill}~\cite{Ryan2010Robust}, and the point-to-point \textsc{waltz}~\cite{Shaka1983Improved} sequence, designed for transfer between the poles of the Bloch sphere. The sequences last from three to five times longer than a Rabi $\pi$ pulse, but all give higher fidelities and greater detuning tolerances. A Bloch sphere representation of the \textsc{waltz} sequence, as compared with a Rabi $\pi$ pulse, is shown for a non-zero detuning in Figure~\ref{WaltzBloch}. The improvement in fidelity afforded by the \textsc{waltz} pulse is visually apparent from the reduced distance of its resultant state from the south pole, as compared with that of the Rabi $\pi$ pulse.

To characterise each inversion sequence experimentally, we measure the ensemble mean fidelity, equal to the normalized population $|c_2|^2$ of state $|2\rangle$. These are shown over a range of normalised laser detunings $\delta/\Omega_{\textnormal{eff}}$ for opposite circular Raman polarizations in Figure~\ref{SpectralScans}, and for orthogonal linear polarizations in Figure~\ref{SpectralScansLinLin}. The displacement of the peak from $\delta=0$ shows the light shift in each case.

\begin{figure}[t!]
  \centering
    \includegraphics[width=8.1cm]{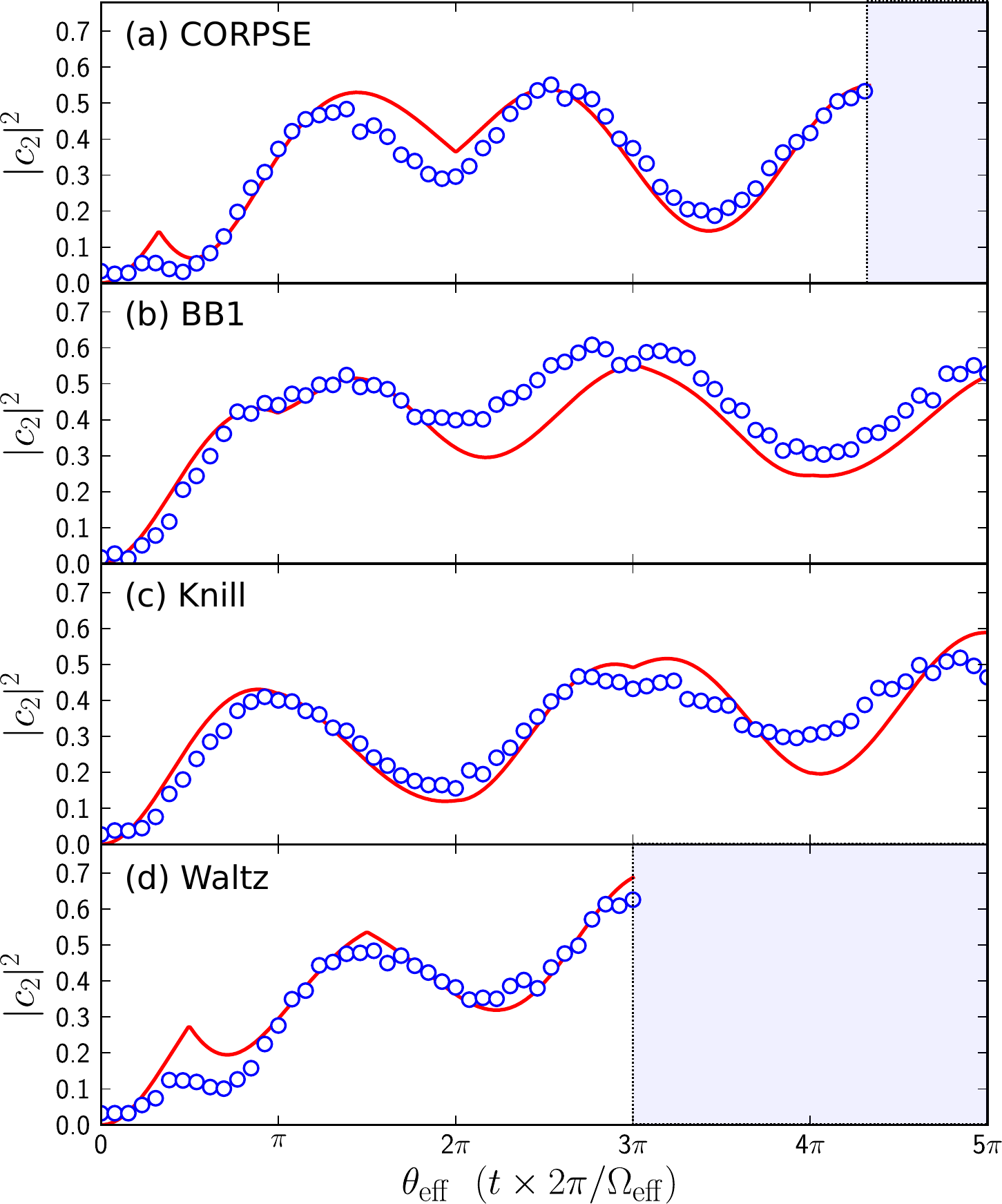}
    \caption{Measured upper state populations (circles) after various $\sigma^{+}-\sigma^{+}$ inversion sequences, as functions of inter-pulse interaction time. Simulations (lines) are again for parameters that reproduce the results for a basic $\pi$ ($180_0$) pulse.}
  \label{TemporalScans}
\end{figure}

By truncating each sequence, we also determine the state population evolution, shown in Figure~\ref{TemporalScans} for circular beam polarizations at the optimum Raman detuning. Experimental fidelities are all presented without correction for the beam overlap factor $S$, described in the appendix.

\section{DISCUSSION}
Our experimental results and theoretical simulations demonstrate general characteristics of coherent manipulations, as well as the differences between different composite pulse sequences. In each case, the single-photon light shift due to the Raman beams is apparent in a detuning of the spectral peak from the low intensity resonance frequency; and features that are resolved in the case of $\pi^{+}-\pi^{-}$ Raman polarizations, for which the light shift is independent of Zeeman sub-state, are blurred into a smooth curve for $\sigma^{+}-\sigma^{+}$ polarizations. Temporal light shift variations as the pulse sections begin and end will distort the composite sequences, but appear to have little effect upon the overall performance. Spatial beam inhomogeneities, the sub-state-dependent Raman coupling strengths, and the Doppler shift distribution, should all to an extent be corrected by the composite pulses.

\begin{figure}[t!]
  \centering
    \includegraphics[width=8.1cm]{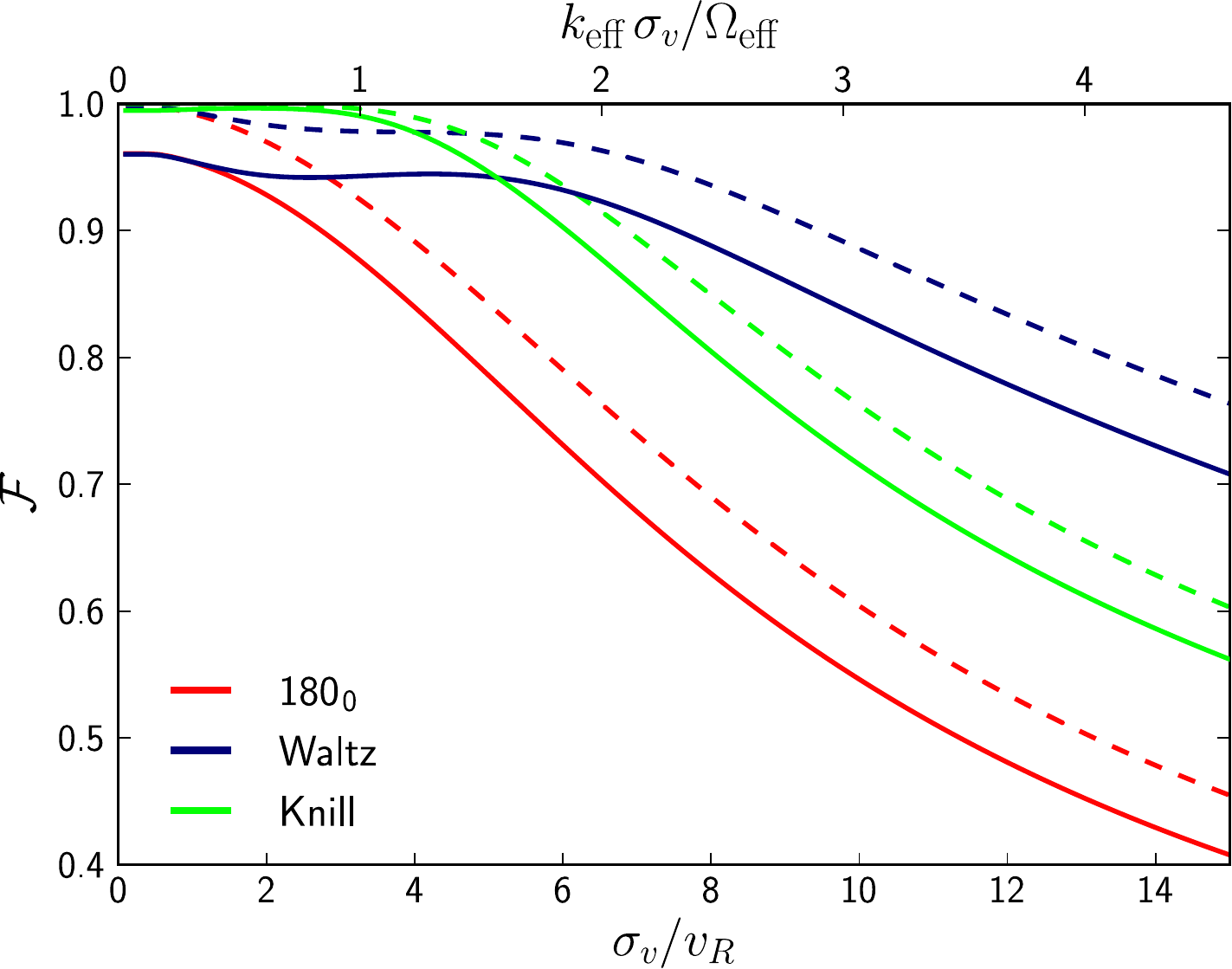}
    \caption{(Color online) Predicted fidelity achievable using \textsc{waltz} (blue/dark-grey) and \textsc{knill} (green/light-grey) composite pulses, compared with a simple $\pi$-pulse (red/mid-grey), for varying velocity distribution widths $\sigma_{v}$ in units of the two-photon recoil velocity $v_{R}$. Dashed lines show predicted behaviour for spin-polarised atoms populating only the $m_{F}=0$ state, while solid lines are for an even distribution across all states $m_{F}=-2\ldots+2$.}
  \label{VelDistChar}
\end{figure}

Our composite pulse sequences vary in the degree to which they cancel \emph{pulse length} and \emph{off-resonance} errors, with the \textsc{corpse} pulse suppressing only off-resonance effects, the \textsc{bb1} tolerating only pulse length errors, and the \textsc{knill} pulse correcting the quadratic terms in both. Accordingly, the \textsc{corpse} sequence shows the greatest insensitivity to detuning, while the \textsc{bb1} and \textsc{knill} pulses show higher peak fidelities. Although the \textsc{bb1} is regarded as the most effective for combating pulse-length errors, we find that pulses that nominally correct for off-resonance effects only can provide greater enhancements in the peak fidelity and spectral width overall.

All three general rotors are out-performed in peak fidelity by the point-to-point \textsc{waltz} sequence, which has already been used for atom interferometer augmentation pulses~\cite{Butts2013Efficient}. This pulse is expected to enhance very small errors, but limit their effect to 5\% for $|f|\lesssim1.1$. In the $\sigma^{+}-\sigma^{+}$ configuration, we observe that the \textsc{waltz} pulse nearly halves the infidelity $\mathcal{I}$ upon which the interferometer contrast depends, from 0.58 for the basic $\pi$ pulse to 0.33. In the $\pi^{+}-\pi^{-}$ configuration, the improvement is from $\mathcal{I} = 0.35$ to $\mathcal{I} = 0.24$, and the fidelity is maintained as predicted for detunings $|\delta_{L}|\lesssim\Omega_{\mathrm{eff}}$; beyond this, it falls more gently so that, at $|\delta_{L}|\approx3\Omega_{\mathrm{eff}}$, it is over five times that for a $\pi$ pulse.

We note that, as the pulse durations in our experiments were chosen by optimizing the $\pi$ pulse fidelity, slight improvements might be possible for the other sequences, both because it is the atomic ensemble average that matters and because the bandwidths of our modulators cause small distortions around the pulse transients.

The close agreement of our experimental results and theoretical simulations demonstrates both the validity of our simple model of the velocity and Zeeman state distributions and the durable underlying coherence of individual atomic states. In each case, the simulation parameters are based upon the measured laser intensities and detunings, which are adjusted within known uncertainties to match the results for a simple $\pi$ pulse under the same experimental conditions; the deduced values are listed in Table~\ref{SimTable}. As the measured efficiencies depend upon experimental conditions that vary between data sets, we have simulated the performance of the $\pi$ pulse and \textsc{knill} and \textsc{waltz} sequences under consistent conditions, for a range of velocity distributions and for two different Zeeman sub-state distributions. Figure~\ref{VelDistChar} demonstrates the expected decrease in fidelity with increasing atom cloud temperature and with the population of multiple Zeeman levels. For low temperatures ($\sigma_{v}\lesssim 5 v_{R}$), the spectral width of the Rabi $\pi$ pulse at $\Omega_{\textnormal{eff}}\sim2\pi\times 350$\,kHz exceeds the Doppler-broadened linewidth, and the peak fidelity is determined by the variation in Raman coupling strength between different Zeeman sub-states; the best fidelity is hence obtained with the superior \emph{pulse-length} error performance of the \textsc{knill} sequence. For warmer samples, Doppler \emph{off-resonance} errors dominate, and the \textsc{waltz} pulse is better. For atoms that are spin-polarized into a single Zeeman level, the performance of all pulses is improved, and the preference for the \textsc{waltz} pulse extends to slightly lower temperatures. Table~\ref{CPTable} summarizes these results for conditions that are typical for our experiments, and also shows the theoretical performance of some other popular composite pulse sequences.

\section{CONCLUSION}
Our results show that the principal errors in the coherent manipulation of cold atoms are due to systematic inhomogeneities in the laser intensity, atomic velocity and Zeeman sub-state, and may hence be significantly reduced by composite pulse techniques. Near the recoil limit, we predict that instead of the the maximum $\pi$ pulse fidelity of 0.96 it should be possible to achieve fidelities in excess of 0.99, allowing many more augmentation pulses to impart a greater separation between the interferometer paths and hence an elevated interferometric sensitivity without losing atoms through spin squeezing. The greater tolerance of Doppler shifts similarly allows interferometry to be performed without further loss through velocity selection.

Atom interferometers require not only augmentation $\pi$ pulses, but beam-splitter/recombiner $\pi/2$ pulses ($90_{0}$), and for these it is likely that quite different composite pulse sequences will be required to minimize the effects of experimental variations upon the composition and phase of the quantum superposition: the solution depends upon the balance of different sources of error, and the relative importance of their different effects upon the final states. Cold atom interferometers are likely to differ in both respects from the NMR systems for which most established composite pulse sequences were developed. In our system, \emph{pulse length} and \emph{off-resonance} errors are not only conflated, they are to some extent correlated, for the light shift is responsible for both.

The best solutions need not be those which optimize the fidelities of the individual interferometer operations, for it is likely that errors after the first beamsplitter, for example, could to some extent be corrected by the recombiner. Indeed, the \textsc{waltz} pulse was developed for decoupling sequences where long chains of $\pi$-pulses, rather than single isolated pulses, are the norm: although, for an equal superposition initial state $|\psi\rangle=\frac{1}{\sqrt{2}}(|g\rangle+e^{i\phi}|e\rangle)$ in the presence of off-resonance errors, a single \textsc{waltz} inversion is worse than a basic $\pi$ pulse, a sequence of \emph{two} \textsc{waltz} pulses gives an efficient $2\pi$ rotation, and when \emph{two} pairs of \textsc{waltz} pulses were applied as in the large area interferometer in \cite{Butts2013Efficient}, readout contrast was increased. The development of composite pulse sequences for atom interferometry should therefore consider the performance of the interferometer as a whole.

\section*{ACKNOWLEDGMENTS}
This work was supported by the UK Engineering and Physical Sciences Research Council (Grant No. GR/S71132/01). RLG acknowledges the use of the IRIDIS High Performance Computing Facility at the University of Southampton. AD acknowledges the use of the QuTiP \cite{Johansson2013qutip} Python toolbox for Bloch sphere illustrations.

\appendix
\renewcommand\appendixname{APPENDIX}
\section{EXPERIMENTAL DETAILS}\label{ExpSet}
$^{85}$Rb atoms are initially trapped and cooled to ${\sim250\,\mu}$K in a standard 3D magneto-optical trap (MOT) to give about $2\times10^7$  atoms in a cloud about $500\,\mu$m in diameter. The MOT magnetic fields are extinguished, the beam intensities ramped down, and the cloud left to thermalise in the 3D molasses for 6\,ms, after which the temperature has fallen to $\sim 50\, \mu$K. The velocity distribution exhibits a double-Gaussian shape, as shown in Figure~\ref{VeloDist}, because atoms at the centre of the molasses undergo more sub-Doppler cooling than those at the edges~\cite{Townsend1995Phasespace}. The $\lvert5S_{1/2},F=2\rangle \rightarrow \lvert5P_{3/2},F=3\rangle$ repumping beam is then extinguished, and the atoms are optically pumped for $300\, \mu s$ into the $\lvert5S_{1/2},F=2\rangle$ ground hyperfine state by the cooling laser, which is detuned to the red of the $\lvert5S_{1/2},F=3\rangle \rightarrow \lvert5P_{3/2},F=4\rangle$ transition. Three mutually orthogonal pairs of shim coils cancel the residual magnetic field at the cloud position, and are calibrated by minimising the spectral width of a Zeeman-split, velocity-insensitive (co-propagating) Raman transition. From the spectral purity of the measured velocity distribution, we deduce the residual magnetic field to be less than 10~mG, equivalent to a Zeeman splitting of $\sim5\,\mathrm{kHz} \times m_{F}$, and hence that the Zeeman sub-levels for each hyperfine state are degenerate to within a fraction of the typical Rabi frequency $\Omega_{\mathrm{eff}}>2\pi \times 200\,\mathrm{kHz}$. After preparation, we apply the Raman pulses to couple the states $\lvert5S_{1/2},F=2\rangle$ and $\lvert5S_{1/2},F=3\rangle$, and then measure the resultant $\lvert5S_{1/2},F=3\rangle$ state population by detecting fluorescence when pumped by the cooling laser.

\begin{figure}[t!]
 \centering
 \includegraphics[width=8.5cm]{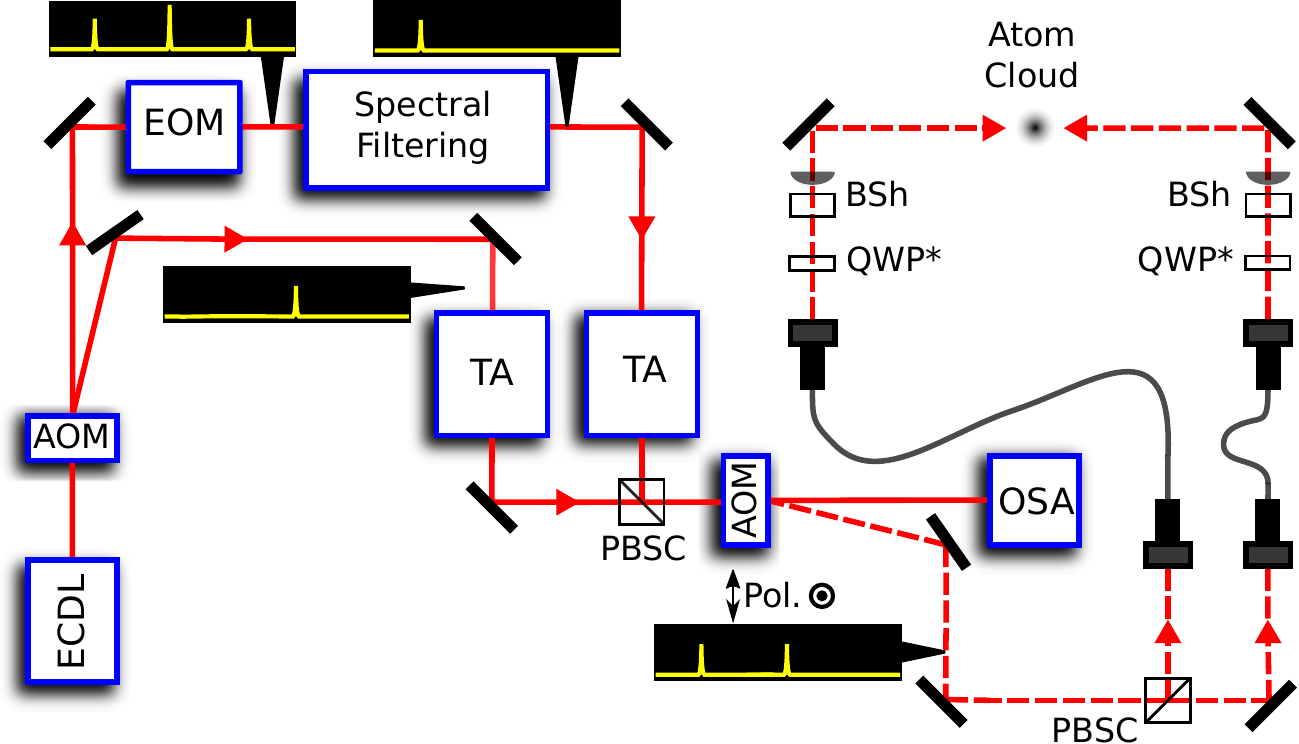}
 \caption{Schematic of the experimental setup of the Raman beams: ECDL -- external-cavity diode laser; PBSC -- polarising beamsplitter cube; TA -- tapered amplifier; OSA -- optical spectrum analyser; BSh -- beam shaper and focussing lens; QWP -- quarter waveplate. *Required only for $\sigma^+-\sigma^+$ Raman transitions. The annotation bubbles show sketches of the beam spectrum at each preparation stage. For clarity, the injection-locked pre-amplifier following the EOM is omitted.\label{OpticalSetup}}
\end{figure}

The apparatus used to generate our Raman pulses is shown schematically in Figure~\ref{OpticalSetup}. The beams are generated by spatially and spectrally splitting the continuous-wave beam from a 780~nm external cavity diode laser, red-detuned from single-photon resonance by $\Delta\approx2\pi\times10$\,GHz. The beam is spatially divided by a 310~MHz acousto-optical modulator (AOM), and the remainder of the microwave frequency shift is generated by passing the undeflected beam from the AOM through a 2.726~GHz electro-optical modulator (EOM). We modulate the EOM phase and frequency using an in-phase and quadrature-phase (IQ) modulator, fed from a pair of arbitrary waveform generators. The carrier wave at the output of the EOM is removed using a polarising beamsplitter cube~\cite{Cooper2012Actively}, and temperature-dependent birefringence within the EOM is countered by active feedback to a liquid crystal phase retarder~\cite{Bateman2010HanschCouillaud}. The remaining off-resonant sideband is removed using a stabilized fibre-optic Mach--Zehnder interferometer~\cite{Cooper2013A}.

Following pre-amplification of the EOM sideband by injection-locking a c.w.\,diode laser, the two spatially separate, spectrally pure Raman beams are then individually amplified by tapered laser diodes, recombined with orthogonal polarisations and passed through an AOM, whose first-order output forms the Raman pulse beam. The AOM rise and fall times alter the effective pulse timing but are not included in the sequence design; proper compensation could further improve the observed fidelity.

Each beam is passed through a Topag GTH-4-2.2 refractive beam shaper and 750~mm focal length lens to produce an approximately square, uniform beam whose intensity varies by only 13\% across the extent of the MOT cloud. The beams measure $\sim2$~mm square and each has an optical power of 50~mW, corresponding to an intensity of $\sim1.3\, \mathrm{W\,cm}^{-2}$. Compared with the large-waist Gaussian beams required for the same spatial homogeneity, this provides a significantly higher intensity and as a result our system exhibits two-photon Rabi frequencies of $\Omega_{\textnormal{eff}}\approx2\pi\times250$\,kHz. Using a shorter focal-length lens in the beam path to produce a smaller top-hat, we have observed a higher Rabi frequency of $\Omega_{\textnormal{eff}}\approx2\pi\times500$\,kHz, but are then limited by the number of atoms that remain within the beam cross-section. Although the phase profile of the top-hat beam is non-uniform~\cite{Boutu2011Highorderharmonic}, we calculate that an individual atom will not traverse a significant phase gradient during a few-$\mu$s pulse sequence.

Because our Raman beams illuminate a smaller region than the cooling and repump beams used to determine the final state population, a fraction of the expanding atom cloud contributes to the normalization signal without experiencing the Raman pulse sequence. We have characterized the time dependence of this effect, and scale our simulated upper state populations in Figures~\ref{SpectralScans}--\ref{TemporalScans} by a factor $S$ as summarized in Table~\ref{SimTable}.

In order to provide a good comparison in the presence of drift (applicable in figures \ref{RotaryEchoes1} \ref{SpectralScans}), we take data on the composite pulses immediately after taking data on its corresponding basic pulses. Experiment shots took 0.4\,s and were repeated 16 times and averaged to improve the signal-to-noise ratio. When taking a spectral scan (figure \ref{SpectralScans}) we ran through different values of the detuning in a pseudo-random sequence to minimise the effects of drift. Data in the temporal scans (figures \ref{RotaryEchoes1} and \ref{TemporalScans}) was not sampled at pseudo-random values of $t$, and therefore the fringe contrast in these plots is subject to drifts in $\Omega_\textnormal{eff}$ from $t=0\rightarrow t_\textnormal{max}$.

\section{THEORETICAL MODEL}\label{TheMod}
We drive two-photon stimulated Raman transitions between the two hyperfine ground states in $^{85}$Rb, as shown in Figure~\ref{RamanDiagram1}. When the two Raman beams with angular frequencies $\omega_{L1,L2}$ and wavevectors $\mathbf{k}_{1,2}$ travel in opposite directions ($\mathbf{k}_{1}\approx-\mathbf{k}_{2}$), the Raman interaction is velocity-sensitive and each transition is accompanied by a two-photon recoil of the atom $\mathbf{k}_{\mathrm{eff}} = \mathbf{k}_{1,2}-\mathbf{k}_{2,1} \approx \pm 2\mathbf{k}_{1,2}$ as a photon is scattered from one Raman beam to the other. The internal state of the atom is therefore mapped to its quantised external momentum state. If an atom is prepared in the lower hyperfine state $\lvert5S_{1/2},F=2\rangle$, which we label $\rvert 1\rangle$, and the Raman transition couples this, via an intermediate virtual state $\lvert 3 \rangle$, to the upper hyperfine state $\lvert5S_{1/2},F=3\rangle$, labelled $\rvert 2\rangle$, then the \emph{momentum-inclusive} basis in which we work is $\rvert 1,\mathbf{p}\rangle$, $\rvert 2,\mathbf{p}+\hbar \mathbf{k}_{\mathrm{eff}}\rangle$. For clarity, we henceforth omit the momenta and leave these implicit in our notation.

The Hamiltonian for the Raman system is~\cite{Young1997Precision}
\begin{equation}
 \displaystyle
  \hat{H} = \frac{\hat{\mathbf{p}}^2}{2m} + \hbar\omega_1\rvert 1\rangle \langle 1\rvert + \hbar\omega_2\rvert 2\rangle \langle 2\rvert + \hbar\omega_3\rvert 3\rangle \langle 3\rvert - \hat{\mathbf{d}}\cdot\mathbf{E}
\end{equation}
where $\hat{\mathbf{p}}$ is the momentum operator and differs in this equation from the ground state momentum $\mathbf{p}$ principally through the introduction of the small Doppler shift to the resonance frequency due to the impulse imparted by the transition. The initial and final electronic states are taken to have energies $\hbar \omega_{1}$ and $\hbar \omega_{2}$, $\hat{\mathbf{d}}$ is the Raman electric dipole operator, acting via all intermediate states $\lvert i\rangle$, and the electric field of the two Raman beams counter-propagating along the $z$ axis is given by
\begin{equation}
 \displaystyle
  \mathbf{E} = \mathbf{E_1} e^{\left(\mathbf{k_1}\cdot\mathbf{z} + \omega_{L1} t + \phi_1\right)} + \mathbf{E_2} e^{\left(\mathbf{k_2}\cdot\mathbf{z} + \omega_{L2} t + \phi_2\right)},
\end{equation}
where on resonance $\omega_{L1} = \omega_3 - \omega_1$ and $\omega_{L2} = \omega_3 - \omega_2$, $\mathbf{E_{1,2}}$ are the Raman beam amplitudes, and we define $\phi = \phi_1 - \phi_2$ as the effective phase of the Raman field. For the analytical solutions to the time-dependent Schr\"{o}dinger equation for this system in the interaction picture, we refer the reader to \cite{Young1997Precision}.

\begin{table*}[t!]
 \begin{ruledtabular}
  \begin{tabular}{l c c c c c c c c c c c}
                                 & Polarization            & $I_{1}$     & $I_{2}$     & $\Delta$ & $B_{z}$ & $\sigma_{1}$ & $\sigma_{2}$ & $a_{1}/a_{2}$ & $S$ & $\Omega_{\mathrm{eff}}$& $t_{\pi}$ \\
                                 &                         & kW~m$^{-2}$ & kW~m$^{-2}$ & $2\pi\times\,$GHz & mG & $v_{R}$ & $v_{R}$ & & & $2\pi\times$kHz & $\mu$s \\ \hline
  Table~\ref{CPTable}              & $\sigma^{+}-\sigma^{+}$ & 12.1 & 12.1 & 12.3 &      & 5   & 22.5 & 4 &      & 250 & 2    \\
  Figure~\ref{VeloDist}            & $\sigma^{+}-\sigma^{+}$ & 3    & 4.6  & 10   & -11  & 1.8 & 7.5  & 4 &      & 4.5 & 110  \\
  Figure~\ref{RotaryEchoes1}       & $\sigma^{+}-\sigma^{+}$ & 12   & 17   & 15   & -101 & 3   & 10   & 3 & 0.95 & 200 & 2.5  \\
  Figure~\ref{SpectralScans}       & $\sigma^{+}-\sigma^{+}$ & 14   & 21   & 9.0  & -11  & 2.5 & 9    & 2 & 0.9  & 357 & 1.4  \\
  Figure~\ref{SpectralScansLinLin} & $\pi^{+}-\pi^{-}$       & 14   & 21   & 8.0  & -11  & 2.5 & 9    & 2 & 0.9  & 417 & 1.2  \\
  Figure~\ref{TemporalScans}       & $\sigma^{+}-\sigma^{+}$ & 14   & 21   & 8.5  & -11  & 2.5 & 9    & 2 & 0.9  & 385 & 1.3  \\
  Figure~\ref{VelDistChar}         & \hphantom{~$^{\dagger}$}$\sigma^{+}-\sigma^{+}$~$^{\dagger}$ & 14   & 21   & 9    & -261 & -    & various    & 0 & 1    & 350$^{\mbox{\textdaggerdbl}}$ & 1.58$^{\mbox{\textdaggerdbl}}$ \\
  \end{tabular}
 \end{ruledtabular}
 \caption{Parameters used in theoretical simulations. In each case, the hyperfine state $m_{F}$ levels are assumed to be equally populated, and the Raman beam intensities --- which may differ because of different beam widths --- are assumed to be spatially and temporally invariant. The velocity distribution is taken to have two Gaussian components with widths $\sigma_{1,2}$, given in units of the single photon recoil velocity $v_{\mathrm{rec}}$, and relative amplitudes $a_{1,2}$; a Gaussian velocity distribution width of $\sigma_{v}$ corresponds to a temperature of $T = m\sigma_{v}^{2}/k_{\mathrm{B}} \equiv 1.48(\sigma_{v}/v_{\mathrm{rec}})^{2}\, \mu\mathrm{K}$. $\dagger$~Also represents $\pi^{+}-\pi^{-}$ as the B-field here is set to compensate for Zeeman-like light shift. \textdaggerdbl~For the single sub-state $m_{F}=0$, $\Omega_{\mathrm{eff}} = 2\pi\times 316$~kHz, $t_{\pi} = 1.43\,\mu$s.}\label{SimTable}
\end{table*}

The effective Rabi frequency of the Raman transition ($\Omega_\textnormal{r}^\prime$ in \cite{Young1997Precision}) depends on (a) the respective Clebsch-Gordan coefficients of the Raman route whose relative amplitudes are shown in Figure~\ref{RamanDiagram1}, (b) the intensity of the driving field, which in our simulations is taken to be temporally `square' and (except in Figure~\ref{RotaryEchoes1}) spatially homogeneous, and (c) the detuning of the driving field from the atomic resonance, which is Doppler-shifted by the atom's motion. It follows that for a Doppler-broadened ensemble of atoms distributed across degenerate sublevels, we expect a distribution of $\Omega_{r}^\prime$ values and therefore a dephasing of atomic states during a Raman pulse. Consequently, the $\pi$ pulse efficiency will be unavoidably limited to much less than unity in the absence of effective error correction.

To simulate the system we numerically calculate the hyperfine state amplitudes $c_{\rvert 1,\mathbf{p}\rangle}(t)$ and $c_{\rvert 2,\mathbf{p}+\hbar\mathbf{k}_{\mathrm{eff}}\rangle}(t)$ for a period $t$ of interaction with the Raman beams, and integrate over all Raman routes and velocity classes. The atoms are taken initially to be evenly distributed across the Zeeman $m_{F}$ sub-levels of $\lvert5S_{1/2},F=2\rangle$, and opposite-circularly polarised Raman beams are considered to drive $\sigma^+$ dipole-allowed transitions via the Raman routes shown in Figure~\ref{RamanDiagram1} where, regardless of the quantisation axis, conservation of angular momentum requires that $\Delta m_F = 0$.

The primary free parameters in our simulations are the velocity distribution, which we model as having two Gaussian components similar to those fitted to the measured distribution in Figure~\ref{VeloDist}, the sampling factor $S$ and the laser intensity $I$. To account for experimental variations, we allow small adjustments from measured values to give a closer fit to the data; the values used for our various simulations are listed in Table~\ref{SimTable}.

\newpage
\providecommand{\noopsort}[1]{}\providecommand{\singleletter}[1]{#1}%

\end{document}